# Title: Gate-controlled BCS-BEC crossover in a two-dimensional superconductor

**Authors:** Yuji Nakagawa[1,2], Yuichi Kasahara[3], Takuya Nomoto[1], Ryotaro Arita[1,4],

Tsutomu Nojima[5], Yoshihiro Iwasa[1,2,4]*

**Affiliations:**

[1]Department of Applied Physics, The University of Tokyo, Hongo 7-3-1, Bunkyo-ku, Tokyo 113-8656, Japan.

[2]Quantum-Phase Electronics Center (QPEC), The University of Tokyo, Hongo 7-3-1, Bunkyo-ku, Tokyo 113-8656, Japan.

[3]Department of Physics, Kyoto University, Kitashirakawa Oiwakecho, Sakyo-ku, Kyoto 606-8502, Japan

[4]RIKEN Center for Emergent Matter Science (CEMS), Hirosawa 2-1, Wako, Saitama 351-0198, Japan

[5]Institute for Materials Research, Tohoku University, Katahira 2-1-1, Aoba-ku, Sendai, 980-0812, Japan

*Correspondence to: iwasa@ap.t.u-tokyo.ac.jp

**Abstract:**

The Bardeen-Cooper-Schrieffer (BCS) condensation and the Bose-Einstein condensation (BEC) are the two extreme limits of the ground state of the paired fermion systems. We report crossover behavior from the BCS condensation to the BEC realized in the two-dimensional (2D) superconductor, electron doped layered material ZrNCl. The phase diagram, established by simultaneous experiments of resistivity and tunneling spectra under the ionic gating, demonstrates the pseudogap phase at the low doping regime. In the low carrier density limit, $T_{BKT}$ (Berezinskii-Kosterlitz-Thouless transition temperature for 2D superconductors) scales as $T_{BKT}/T_F = 0.12$, where $T_F$ is the Fermi temperature, which is consistent with the theoretical upper bound expected in the BCS-BEC crossover regime. The present results indicate that the gate-doped semiconductor provides an ideal platform for the 2D BCS-BEC crossover without any added complexity, such as magnetic orders and density waves.

**One Sentence Summary:**

Two-dimensional BCS-BEC crossover is demonstrated in the Li intercalated ZrNCl superconductors.

**Main Text:**

The condensation of fermions is fundamental to a variety of systems, ranging from neutron stars in the universe to superconductors in solids and ultracold atomic gases realized in laboratories (*1-7*). This condensation behavior is usually classified into two limits and described by different theories: the Bardeen-Cooper-Schrieffer (BCS) theory and the Bose-Einstein condensation (BEC). The former explains superfluidity in the weak-coupling or high-density limit, where individual fermions directly condense to the coherent state of fermion pairs. This type of condensation is typically observed in superconductivity (SC) of electrons. The latter occurs in the strong-coupling, low-density limit. Fermions first form pairs that behave as bosons, and these bosons finally undergo the BEC to the superfluid state (*1*). The archetype of this phenomenon is seen in fermionic gases (*3*). Between these two limits, theories do not expect a phase transition. They are continuously connected through an intermediate regime, exhibiting universal behavior called the BCS-BEC crossover (*2, 7*).

The experimental observation of how the BCS-BEC crossover occurs is a significant challenge and crucial for constructing a unified understanding of fermionic condensation. Promising candidate systems are ultracold atomic gases and superconductors because the coupling strengths can be controlled in a quasi-continuous manner. In the ultracold atomic gasses, the coupling strength is highly modulated via the Feshbach resonance (*7*), which has brought the system into the crossover regime from the BEC limit (*4, 6*). The approach from the BCS limit is expected in superconductors by controlling the carrier density and, consequently, the coupling strength.

In superconductors, the dimensionless coupling strength is determined by $\Delta/E_F$, where $\Delta$ is the superconducting gap, and $E_F$ is the Fermi energy measured from the bottom of the conduction band. As $\Delta/E_F$ is increased by enhanced pairing interaction or reduced carrier density, the system enters the BCS-BEC crossover regime accompanied by the enhancement of $T_c/T_F$, where $T_c$ and $T_F$ are the superconducting critical temperature and the Fermi temperature, respectively. As represented in the well-known Uemura plot (*8*), conventional metallic superconductors, such as Nb and Al, reside deep inside the BCS limit ($T_c/T_F \ll 1$), whereas the "exotic" superconductors, including cuprates, organics, heavy fermions, and iron-based superconductors, locate rather close to the BCS-BEC crossover region with similar $T_c/T_F$ values. However, even with these systems, the coupling strengths are not high enough to reach the BEC limit beyond the crossover regime. Also, in these systems, the low carrier density limit mostly exhibits strong electron correlation effects with magnetic ordering (*9*), clouding the crossover phenomena by added complexity. Therefore, the clear demonstration of the BCS-BEC crossover is still a key challenge of SC. Although FeSe systems (*10, 11*), magic-angle twisted bilayer graphene (*12*), and layered nitrides (*13*) have been studied in this context, the wide-range control of $\Delta/E_F$, i.e., venturing from the BCS regime to the BEC limit, has not been realized to date.

The superconductor we studied is a lithium-intercalated layered nitride, Li$_x$ZrNCl (*13, 14*) (Fig. 1A). Lithium supplies electrons to the double honeycomb ZrN layer, which itself is a band insulator at $x = 0$. A single conduction band originating from Zr 4$d$ orbitals hybridized with N 2$p$ orbitals exists at each corner of the hexagonal Brillouin zone (K and K' points). A unique property reported in polycrystalline bulk samples is the $T_c$ enhancement toward low doping regimes until SC is quenched by disorder (*15*). Further increase of $T_c$ above 15 K is theoretically proposed to exist in the low-carrier-density limit (*16, 17*), which leads to a substantial increase of $\Delta/E_F$ and the BCS-BEC crossover.

On the other hand, the single-crystal measurement of pristine ZrNCl has been performed by using ionic gating methods (*18, 19*). The interface SC was realized in electrostatic electric double layer transistors (EDLTs), where the intrinsic two-dimensional (2D) natures of SC have been elucidated in this 2D material. Recently, we introduced a modified device structure designed for an electrochemical intercalation mode as well as the tunneling spectroscopy measurements (*13*). The electrical control of the Li amount in the single crystals has been proven to be more effective in pursuing the low carrier density limit of Li$_x$ZrNCl, as well as Li$_x$HfNCl, than the conventional chemical intercalation methods in polycrystalline samples using *n*-BuLi.

Here, we report the superconducting behavior of Li$_x$ZrNCl in the even lower carrier density regime down to $x = 0.0038$. With diluting the carrier density, $T_c$ shows the maximum value of 19.0 K at $x = 0.011$. $\Delta/E_F$ increases above 0.3, which is considered to be the border between the weak-coupling BCS limit and the BCS-BEC crossover regime in 2D systems (*20*). The Berezinskii-Kosterlitz-Thouless (BKT) transition temperature $T_{BKT}$ reaches $0.12T_F$, which is close to the maximum value generally predicted in the fermion systems, especially in the BCS-BEC crossover regime (*21, 22*). More importantly, the tunneling spectroscopy simultaneously performed with the resistivity measurements revealed that the pseudogap state, in which the fermions form pairs without condensation (*6, 7*), is enlarged in the diluted regime. These results are an unprecedented and unambiguous demonstration of a crossover from the BCS to BEC regime of a 2D superconductor by scanning the doping level by nearly two orders of magnitude from $x = 0.28$ to $0.0038$.

Figure 1B presents a schematic of our ionic-gating device structure. In addition to the Hall bar structure, narrow electrodes for the tunneling spectroscopy were prepared, followed by covering the device with the PMMA resist. The cover outside of the channel region was removed such that Li ions in the electrolyte could intercalate from both exposed sides of the flake. During the application of the gate voltage $V_G$, we traced the intercalation process through the measurement of the source-drain current (Fig. 1C). The forward scan shows a sharp increase of current corresponding to the sudden intercalation, possibly because of the shift of the layers causing a change in the stacking pattern (*14*). In contrast, the deintercalation in the reverse scan is rather gradual. This process enabled us to control the intercalation level down to the lowest level of $x = 0.0038$, which has never been achieved so far. The Hall effect showed a systematic change depending on $V_G$ (Fig. 1D) and was almost independent of temperature and measurement position on the flake (Fig. S2, S3). Here, we determined the Li concentration $x$ from the Hall carrier density at 150 K, assuming that one electron is supplied to the ZrNCl conduction layer per Li ion.

In Fig. 2A and B, we show the temperature dependence of resistivity at various doping levels. For the SC, $T_c$ increases from 11.5 K to 19.0 K by reducing the doping level. We defined $T_c$ at the midpoint of the resistive transition. $T_c$ reached 19.0 K, which is higher than the previously reported value (15.4 K) (*13, 15*), marking a record high value for ZrNCl systems. Interestingly, this is realized by reducing the doping level. Below $x = 0.011$, corresponding to a carrier density of $n_{3D} = 2.1 \times 10^{20}$ cm$^{-3}$, $T_c$ starts to decrease and forms a peak structure in its phase diagram (see Fig. 3E).

The resistive transition in the highly doped regime is sharp, whereas it is substantially broadened in the lightly doped regime (Fig. 2B). As reported previously (*13*), this feature represents a dimensional crossover from anisotropic three-dimensional (3D) to 2D SC because of the reduced interlayer hopping as discussed later. In the 2D regime, the superconducting transition is described by the BKT transition, and the temperature dependence of resistivity follows the Halperin-Nelson equation (*23, 24*).

$$\rho(T) = a\rho_\mathrm{N} \exp\left[-2\left\{\frac{b(T_c' - T)}{T - T_\mathrm{BKT}}\right\}^{1/2}\right], \qquad (1)$$

where $a$, $b$, and $T_c$' are fitting parameters, and $\rho_\mathrm{N}$ is resistivity in the normal state. According to this formula, $(d \ln\rho/d T)^{-2/3}$ is proportional to $(T - T_\mathrm{BKT})$ around $T_\mathrm{BKT}$, by which we determine $T_\mathrm{BKT}$ as the temperature-axis intercept of the linear extrapolation (Fig. 2C, inset). With this $T_\mathrm{BKT}$, Eq. (1) well reproduces the transition (for Fig. 2C, $T_\mathrm{BKT}$ = 17.9 K, $a\rho_\mathrm{N}$ = 1.59 mΩ cm, $b$ = 1.23, and $T_c$' = 20.1 K). The two-dimensionality was also confirmed by the temperature dependence of the in-plane upper critical field, $H_{c2}^\parallel (T) \propto \sqrt{1 - T/T_c}$, which is typical behavior within 2D superconductors explained by the Ginzburg-Landau (GL) model (*13*).

The dimensional crossover from anisotropic 3D to 2D SC caused by the reduction of carrier density is a unique and unexpected phenomenon, particularly when we recall that the out-of-plane lattice parameter decreases with the Li concentration (*15*). This feature should be attributed to the peculiar nature of rhombohedral stacking of ZrNCl layers, where the unit cell is composed of 3 layers (Fig. 1A). According to the symmetry argument, the interlayer hopping at the K point of the hexagonal Brillouin zone becomes exactly zero up to the second nearest layer in the rhombohedral structure (*25*). Our density functional theory (DFT) calculation with improved energy resolution also confirms the indiscernible *K-H* dispersion (see Supplementary Materials). When the doping level is reduced, the Fermi surface converges to the K-point. Consequently, the interlayer coupling is weakened, and the out-of-plane coherence length is significantly shortened.

Figure 2D shows the temperature dependence of the out-of-plane upper critical field $H_{c2}$ at each doping level (see Fig. S4 for the determination). Based on the GL model, $\mu_0 H_{c2}(T) = \Phi_0/(2\pi\xi^2)\,(1-T/T_c)$ with $\Phi_0$ a flux quantum, we derived the in-plane coherence length at zero-temperature ($\xi$) by using the slope of the $\mu_0 H_{c2}$-$T$ line (Fig. 2E). Even after $T_c$ starts to drop below $x = 0.011$, $\xi$ keeps decreasing, indicating that the strongly coupled small Cooper pairs are realized in the low carrier density regime.

During the cooling process for the transport measurements, we were able to perform tunneling spectroscopy, thanks to the formation of a Schottky barrier at the electrode interface which served as the tunneling barriers (*13*). Figure 3A displays symmetrized tunneling spectra $dI/dV$ at $T = 2$ K (see Fig. S5 for the raw data). We normalized the spectra at 55 K after the subtraction of the channel resistance contribution (*13*). The superconducting gap structure is clearly observed and becomes wider in the low doping regime. The gap energy $\Delta$ is determined by fitting the spectra to the Dynes function (*26*), and its doping dependence is plotted in Fig. 3B. It should be noted that, at the high doping level, the present results agree well with the previously reported values determined by specific heat measurements on bulk polycrystalline samples (*27*). The value of $2\Delta/k_B T_c$ is around 3.5 at the high doping level, indicating that SC in the highly doped state is within the BCS regime. As carrier density is decreasing, the stronger coupling is realized since $2\Delta/k_B T_c$ increases dramatically, reaching 6.0 at the lowest carrier density.

We show the temperature-evolution of the tunneling spectra at a low-doping level in Fig. 3C. The gap structure smoothly closes at high temperatures. However, even above $T_c$, the gap feature still remains. The inset of Fig. 3C displays the temperature dependence of the zero-bias conductance (ZBC), $dI/dV$ at $V = 0$. We determined the gap-opening temperature $T^*$ by a 1% drop of the ZBC from the value at high temperatures. Figure 3D shows the temperature dependence of $\Delta$ at $x = 0.0066$ and $0.13$ (see Fig. S7 for further information). We set $\Delta = 0$ at temperatures above $T^*$, and the scaled BCS-type gap function was calculated by using $T_c$. For $x = 0.0066$, $T^*$ is more than twice as high as $T_c$, demonstrating the stabilization of the pseudogap state. This is in marked contrast with the case of $x = 0.13$, where $T_c$ and $T^*$ match with each other. Such behavior is summarized in the phase diagram, Fig. 3E, showing that the pseudogap state is significantly developed in the low carrier density regime.

The substantial values of $T_c$ and $\Delta$ in the low carrier density regime give us the required condition to reach the BCS-BEC crossover regime. In Fig. 4A, we show the doping dependence of $\Delta/E_F$ and $1/(k_F\xi)$, where $k_F$ is the Fermi wave vector. The increase of the latter value corresponds to a smaller number of overlapped pairs since $1/k_F$ represents the interparticle distance. $E_F$ and $k_F$ are estimated by the carrier density and the doping-independent effective mass of ZrNCl, $0.9m_0$, which was experimentally derived from the specific heat measurements (*27*). This effective mass is also supported by the DFT calculation combined with electron-phonon interactions (see Supplementary Materials for details). The monotonic increases in both parameters toward the low doping limit represent the continuous shift from the BCS limit toward the BCS-BEC crossover regime. Importantly, $\Delta/E_F$ becomes larger than $0.3$, corresponding to the BCS-BEC crossover region (*20*). $1/(k_F\xi) \sim 0.36$ means that a few Cooper pairs overlap, being distinct from the assumption in the BCS theory.

By taking $\Delta/E_F$ or $1/(k_F\xi)$ as a bottom axis, we depict the BCS-BEC crossover phase diagram (Fig. 4B). Both the superconducting and the pseudogap phase systematically evolve as predicted by theories (*2, 7*). Since the 2D SC is realized in the lightly doped regime, it should be compared to the theory developed for 2D systems, where $T_{BKT}/T_F = 0.125$ is derived as the upper bound for the transition in all fermion systems (*22*). The present observation reached the value of $T_{BKT}/T_F = 0.116$ ($T_c/T_F = 0.121$) at $\Delta/E_F = 0.36$, which is close enough to the theoretical value. This agreement demonstrates that superconductors, where charged electron pairs in the lattice potentials matter, can reach the general upper bound. Figure 4C shows the Uemura plot of our results together with other materials (*11-13, 28-30*). Li$_x$ZrNCl approaches the limit of $T_c/T_F = 0.125$ with decreasing $T_F$ from the BCS limit (lower right side of the plot), providing a strong piece of evidence that the BCS-BEC crossover is indeed reached. The $T_c$ deduction below $x = 0.011$ (Fig. 3E) should be attributed to the limitation of $T_c$ ($T_{BKT}$) scaled by $T_F$ in the BCS-BEC crossover regime rather than to other causes, such as a tendency toward electron localization.

Here, it is fruitful to discuss pseudogap states in several materials and compare them with the present system. In the unconventional superconductors, the pseudogap states are frequently observed close to the insulating phases, but they tend to be complex because of strong electron correlation effects and magnetic ordering in many cases (*9*). In stark contrast, Li$_x$ZrNCl offers a much simpler testbed because the undoped state at $x = 0$ is a band-insulator free from electron correlation effects, magnetic orders, and density waves. Also, unlike magic-angle twisted bilayer graphene, the simple parabolic conduction band does not apply further constraints on $T_c$ (*22*) and makes Li$_x$ZrNCl a model system for the BCS-BEC crossover.

Even in disordered thin films, the pseudogap state is observed due to the localized Cooper pairs (*31*). It is clear that such a localized model is unlikely for the present system because resistivity shows metallic behavior. Furthermore, as summarized in Table S1, the mean free path in the Li$_x$ZrNCl system is comparable to the Pippard coherence length $\xi_{Pippard} = \hbar v_F/\pi\Delta$, where $\hbar$ and $v_F$ are Dirac's constant and the Fermi velocity, respectively, indicating that the system is far from the dirty limit.

These considerations strongly support that the observed pseudogap state in Li$_x$ZrNCl is attributed to the preformed pair formation in the BCS-BEC crossover scenario. We note that the pseudogap state is observable in a wide range, even when $\Delta/E_F$ is less than 0.3. In the bulk study, an NMR measurement on polycrystalline Li$_x$ZrNCl samples reported a pseudogap state on the high doping side of the superconducting dome (*32*). $T^*$ was reported to be ~25 K when $T_c$ = 15 K ($x$ = 0.08), which agrees with the present result. The distinct appearance of pseudogap states is presumably attributed to the two-dimensionality, which prefers the pseudogap formation (*33*). This scenario can be applied to the pseudogap state observed at a LaAlO$_3$/SrTiO$_3$ interface (*34*). This system is still distant from the BCS-BEC crossover regime (Fig. 4C), but its two-dimensionality makes the pseudogap observable.

In conclusion, we have demonstrated a 2D BCS-BEC crossover through the systematic tuning of the coupling strength of SC in Li$_x$ZrNCl. The 2D BCS-BEC crossover was realized in the present system owing to the dimensional crossover from anisotropic 3D to 2D upon reducing the carrier density. This crossover should be compared to the array of 2D clouds of Fermi gases (*35*) because their dimensionality is also affected by the coupling strength. These two fermion systems in different fields advances our understanding of fermion condensation physics, which leads to novel superfluidity at high temperatures.

**References and Notes:**

1. D. M. Eagles, Possible pairing without superconductivity at low carrier concentrations in bulk and thin-film superconducting semiconductors. *Phys. Rev.* **186**, 456-463 (1969).

2. C. A. R. Sá de Melo, M. Randeria, J. R. Engelbrecht, Crossover from BCS to Bose superconductivity: Transition temperature and time-dependent Ginzburg-Landau theory. *Phys. Rev. Lett.* **71**, 3202-3205 (1993).

3. M. Greiner, C. A. Regal, D. S. Jin, Emergence of a molecular Bose–Einstein condensate from a Fermi gas. *Nature* **426**, 537-540 (2003).

4. C. A. Regal, M. Greiner, D. S. Jin, Observation of resonance condensation of fermionic atom pairs. *Phys. Rev. Lett.* **92**, 040403 (2004).

5. A. Gezerlis, J. Carlson, Low-density neutron matter. *Phys. Rev. C* **81**, 025803 (2010).

6. J. P. Gaebler, J. T. Stewart, T. E. Drake, D. S. Jin, A. Perali, P. Pieri, G. C. Strinati, Observation of pseudogap behaviour in a strongly interacting Fermi gas. *Nat. Phys.* **6**, 569-573 (2010).

7. M. Randeria, E. Taylor, Crossover from Bardeen-Cooper-Schrieffer to Bose-Einstein condensation and the unitary Fermi gas. *Annu. Rev. Condens. Matter Phys.* **5**, 209-232 (2014).

8. Y. J. Uemura, L. P. Le, G. M. Luke, B. J. Sternlieb, W. D. Wu, J. H. Brewer, T. M. Riseman, C. L. Seaman, M. B. Maple, M. Ishikawa, D. G. Hinks, J. D. Jorgensen, G. Saito, H. Yamochi, Basic similarities among cuprate, bismuthate, organic, Chevrel-phase, and heavy-fermion superconductors shown by penetration-depth measurements. *Phys. Rev. Lett.* **66**, 2665 (1991).

9. B. Keimer, S. A. Kivelson, M. R. Norman, S. Uchida, J. Zaanen, From quantum matter to high-temperature superconductivity in copper oxides. *Nature* **518**, 179-186 (2015).

10. S. Rinott, K. B. Chashka, A. Ribak, E. D. L. Rienks, A. Taleb-Ibrahimi, P. Le Fevre, F. Bertran, M. Randeria, A. Kanigel, Tuning across the BCS-BEC crossover in the multiband superconductor $Fe_{1+y}Se_xTe_{1-x}$: An angle-resolved photoemission study. *Sci. Adv.* **3**, e1602372 (2017).

11. S. Kasahara, T. Watashige, T. Hanaguri, Y. Kohsaka, T. Yamashita, Y. Shimoyama, Y. Mizukami, R. Endo, H. Ikeda, K. Aoyama, T. Terashima, S. Uji, T. Wolf, H. v. Löhneysen, T. Shibauchi, Y. Matsuda, Field-induced superconducting phase of FeSe in the BCS-BEC cross-over. *Proc. Natl. Acad. Sci. U. S. A.* **111**, 16309-16313 (2014).

12. Y. Cao, V. Fatemi, S. Fang, K. Watanabe, T. Taniguchi, E. Kaxiras, P. Jarillo-Herrero, Unconventional superconductivity in magic-angle graphene superlattices. *Nature* **556,** 43–50 (2018).

13. Y. Nakagawa, Y. Saito, T. Nojima, K. Inumaru, S. Yamanaka, Y. Kasahara, Y. Iwasa, Gate-controlled low carrier density superconductors: Toward the two-dimensional BCS-BEC crossover. *Phys. Rev. B* **98**, 064512 (2018).

14. Y. Kasahara, K. Kuroki, S. Yamanaka, Y. Taguchi, Unconventional superconductivity in electron-doped layered metal nitride halides $M$N$X$ ($M$ = Ti, Zr, Hf; $X$ = Cl, Br, I). *Physica C* **514**, 354–367 (2015).

15. Y. Taguchi, A. Kitora, Y. Iwasa, Increase in $T_c$ upon reduction of doping in $Li_x$ZrNCl superconductors, *Phys. Rev. Lett.* **97**, 107001 (2006).

16. K. Kuroki, Spin-fluctuation-mediated *d+id'* pairing mechanism in doped b-*M*NCl (*M* = Hf, Zr) superconductors. *Phys. Rev. B* **81**, 104502 (2010).

17. M. Calandra, P. Zoccante, F. Mauri, Universal increase in the superconducting critical temperature of two-dimensional semiconductors at low doping by the electron-electron interaction. *Phys. Rev. Lett.* **114**, 077001 (2015).

18. J. T. Ye, Y. J. Zhang, R. Akashi, M. S. Bahramy, R. Arita, Y. Iwasa, Superconducting dome in a gate-tuned band insulator. *Science* **338**, 1193-1196 (2012).

19. Y. Saito, Y. Kasahara, J. T. Ye, Y. Iwasa, T. Nojima, Metallic ground state in an ion-gated two-dimensional superconductor. *Science* **350**, 409-413 (2015).

20. L. He, H. Lü, G. Cao, H. Hu, X. -J. Liu, Quantum fluctuations in the BCS-BEC crossover of two-dimensional Fermi gases. *Phys. Rev. A* **92**, 023620 (2015).

21. S. S. Botelho and C. A. R. Sá de Melo, Vortex-antivortex lattice in ultracold fermionic gases. *Phys. Rev. Lett.* **96**, 040404 (2006).

22. T. Hazra, N. Verma, M. Randeria, Bounds on the superconducting transition temperature: Applications to twisted bilayer graphene and cold atoms. *Phys. Rev. X* **9**, 031049 (2019).

23. B. I. Halperin, D. R. Nelson, Resistive transition in superconducting films. *J. Low Temp. Phys.* **36**, 599-616 (1979).

24. A. M. Kadin, K. Epstein, A. M. Goldman, Renormalization and the Kosterlitz-Thouless transition in a two-dimensional superconductor. *Phys. Rev. B* **27**, 6691-6702 (1983).

25. R. Akashi, Y. Iida, K. Yamamoto, K. Yoshizawa, Interference of the Bloch phase in layered materials with stacking shifts. *Phys. Rev. B* **95**, 245401 (2017).

26. R. C. Dynes, V. Narayanamurti, J. P. Garno, Direct measurement of quasiparticle-lifetime broadening in a strong-coupled superconductor, *Phys. Rev. Lett.* **41**, 1509-1512 (1978).

27. Y. Kasahara, T. Kishiume, T. Takano, K. Kobayashi, E. Matsuoka, H. Onodera, K. Kuroki, Y. Taguchi, Y. Iwasa, Enhancement of pairing interaction and magnetic fluctuations toward a band insulator in an electron-doped Li$_x$ZrNCl superconductor. *Phys. Rev. Lett.* **103**, 077004 (2009).

28. Q. -W. Wang, Z. Li, W. -H. Zhang, Z. -C. Zhang, J. -S. Zhang, W. Li, H. Ding, Y. -B. Ou, P. Deng, K. Chang, J. Wen, C. -L. Song, K. He, J. -F. Jia, S. -H. Ji, Y. -Y. Wang, L. -L. Wang, X. Chen, X. -C. Ma, Q. -K. Xue, Interface-Induced High-Temperature Superconductivity in Single Unit-Cell FeSe Films on SrTiO$_3$. *Chin. Phys. Lett.* **29**, 037402 (2012).

29. Y. J. Uemura, Condensation, excitation, pairing, and superfluid density in high-$T_c$ superconductors: the magnetic resonance mode as a roton analogue and a possible spin-mediated pairing. *J. Phys.: Condens. Matter* **16**, S4515-S4540 (2004).

30. X. Lin, Z. Zhu, B. Fauqué, K. Behnia, Fermi surface of the most dilute superconductor. *Phys. Rev. X* **3**, 021002 (2013).

31. B. Sacépé, T. Dubouchet, C. Chapelier, M. Sanquer, M. Ovadia, D. Shahar, M. Feigel'man, L. Ioffe, Localization of preformed Cooper pairs in disordered superconductors. *Nat. Phys.* **7**, 239-244 (2011).

32. H. Kotegawa, S. Oshiro, Y. Shimizu, H. Tou, Y. Kasahara, T. Kishiume, Y. Taguchi, Y. Iwasa, Strong suppression of coherence effect and appearance of pseudogap in the layered nitride superconductor Li$_x$ZrNCl: $^{91}$Zr- and $^{15}$N-NMR studies. *Phys. Rev. B* **90**, 020503(R) (2014).

33. Q. Chen, I. Kosztin, B. Jankó, K. Levin, Superconducting transitions from the pseudogap state: *d*-wave symmetry, lattice, and low-dimensional effects. *Phys. Rev. B* **59**, 7083-7093 (1999).

34. C. Richter, H. Boschker, W. Dietsche, E. Fillis-Tsirakis, R. Jany, F. Loder, L. F. Kourkoutis, D. A. Muller, J. R. Kirtley, C. W. Schneider, J. Mannhart, Interface superconductor with gap behaviour like a high-temperature superconductor. *Nature* **502**, 528-531 (2013).

35. M. G. Ries, A. N. Wenz, G. Zürn, L. Bayha, I. Boettcher, D. Kedar, P. A. Murthy, M. Neidig, T. Lompe, S. Jochim, Observation of pair condensation in the quasi-2D BEC-BCS crossover. *Phys. Rev. Lett.* **114**, 230401 (2015).

36. S. Yamanaka, H. Kawaji, K. -i. Hotehama, M. Ohashi, A new layer-structured nitride superconductor. Lithium-intercalated β-zirconium nitride chloride, Li$_x$ZrNCl. *Adv. Mater.* **8**, 771-774 (1996).

37. J. K. Hulm, C. K. Jones, D. W. Deis, H. A. Fairbank, P. A. Lawless, Superconducting Interactions in Tin Telluride. *Phys. Rev.* **169**, 388-394 (1968).

38. T. D. Thanh, A. Koma, S. Tanaka, Superconductivity in the BaPb$_{1-x}$Bi$_x$O$_3$ system. *Appl. Phys.* **22**, 205-212 (1980).

39. T. Takano, A. Kitora, Y. Taguchi, Y. Iwasa, Modulation-doped-semiconductorlike behavior manifested in magnetotransport measurements of Li$_x$ZrNCl layered superconductors. *Phys. Rev. B* **77**, 104518 (2008).

40. R. Akashi, M. Ochi, R. Suzuki, S. Bordács, Y. Tokura, Y. Iwasa, R. Arita, Two-Dimensional Valley Electrons and Excitons in Noncentrosymmetric 3*R*-MoS$_2$, *Phys. Rev. Appl.* **4**, 014002 (2015).

41. R. Weht, A. Filippetti, W. E. Pickett, Electron doping in the honeycomb bilayer superconductors (Zr, Hf) NCl. *Europhys. Lett.* **48**, 320-325 (1999).

42. R. Heid, K. P. Bohnen, *Ab Initio* lattice dynamics and electron-phonon coupling in Li$_x$ZrNCl. *Phys. Rev. B* **72**, 134527 (2005).

43. M. Lüders, M. A. L. Marques, N. N. Lathiotakis, A. Floris, G. Profeta, L. Fast, A. Continenza, S. Massidda, E. K. U. Gross, *Phys. Rev. B* **72**, 024545 (2005).

44. M. Kawamura, R. Akashi, S. Tsuneyuki, Anisotropic superconducting gaps in YNi$_2$B$_2$C: A first-principles investigation. *Phys. Rev. B* **95**, 054506 (2017).

45. R. Akashi, K. Nakamura, R. Arita, M. Imada, High-temperature superconductivity in layered nitrides β-Li$_x$*M*NCl (*M* = Ti, Zr, Hf): Insights from density functional theory for superconductors. *Phys. Rev. B* **86**, 054513 (2012).

46. P. Giannozzi, S. Baroni, N. Bonini, M. Calandra, R. Car, C. Cavazzoni, D. Ceresoli, G. L. Chiarotti, M. Cococcioni, I. Dabo, A. Dal Corso, S. de Gironcoli, S. Fabris, G. Fratesi, R. Gebauer, U. Gerstmann, C. Gougoussis, A. Kokalj, M. Lazzeri, L. Martin-Samos, N. Marzari, F. Mauri, R. Mazzarello, S. Paolini, A. Pasquarello, L. Paulatto, C. Sbraccia, S. Scandolo, G. Sclauzero, A. P. Seitsonen, A. Smogunov, P. Umari, R. M. Wentzcovitch, QUANTUM ESPRESSO: a modular and open-source software project for quantum simulations of materials. *J. Phys.: Condens. Matter* **21**, 395502 (2009).

47. P. Giannozzi, O. Andreussi, T. Brumme, O. Bunau, M. B. Nardelli, M. Calandra, R. Car, C. Cavazzoni, D. Ceresoli, M. Cococcioni, N. Colonna, I. Carnimeo, A. Dal Corso, S. de Gironcoli, P. Delugas, R. A. DiStasio Jr., A. Ferretti, A. Floris, G. Fratesi, G. Fugallo, R. Gebauer, U. Gerstmann, F. Giustino, T. Gorni, J. Jia, M. Kawamura, H. -Y. Ko, A. Kokalj, E. Küçükbenli, M. Lazzeri, M. Marsili, N. Marzari, F. Mauri, N. L. Nguyen, H. -V. Nguyen, A. Otero-de-la-Roza, L. Paulatto, S. Poncé, D. Rocca, R. Sabatini, B. Santra, M. Schlipf, A. P. Seitsonen, A. Smogunov, I. Timrov, T. Thonhauser, P. Umari, N. Vast, X. Wu, S. Baroni Advanced capabilities for materials modelling with Quantum ESPRESSO. *J. Phys.: Condens. Matter* **29**, 465901 (2017).

48. D. R. Hamann, Optimized norm-conserving Vanderbilt pseudopotentials. *Phys. Rev. B* **88**, 085117 (2013).

49. P. Scherpelz, M. Govoni, I. Hamada, G. Galli, Implementation and validation of fully relativistic *GW* calculations: Spin–orbit coupling in molecules, nanocrystals, and solids. *J. Chem. Theory Comput.* **12**, 3523-3544 (2016).

50. J. P. Perdew, K. Burke, M. Ernzerhof, Generalized gradient approximation made simple. *Phys. Rev. Lett.* **77**, 3865-3868 (1996).

51. S. Baroni, S. de Gironcoli, A. Dal Corso, P. Giannozzi, Phonons and related crystal properties from density-functional perturbation theory. *Rev. Mod. Phys.* **73**, 515-562 (2001).

**Acknowledgments:** We thank K. Kanoda, T. Shibauchi, T. Hanaguri, M. Nakano, R. Akashi, M. Hirayama, and F. Qin for fruitful discussions. We also thank M. Heyl for critical reading and correction of the manuscript.

**Funding:** This work was supported by A3 Foresight Program and JSPS KAKENHI Grant Numbers JP19H05602 and JP17J08941. Y.N. was supported by the Materials Education program for the future leaders in Research, Industry, and Technology (MERIT).

**Author contributions:** Y. N., T. N. and Y. I. conceived and designed the experiments. Y.N. fabricated the devices, performed measurements and analyzed the data. Y.K. grew the single crystal of ZrNCl. T. N. and R. A. conducted the density functional theory calculations. All authors discussed the results and wrote the manuscript.

**Competing interests:** Authors declare no competing interests.

**Data and materials availability:** The data that support the plots and other findings of this study are available from the corresponding author upon reasonable request.

**Supplementary Materials:**

Materials and Methods

       Device fabrication

       Measurement details

Supplementary Text

1. Low carrier density superconductors
2. Temperature dependence of Hall coefficient
3. Uniformity of intercalation
4. Measurement of the upper critical field
5. Raw data and symmetrization of tunneling spectra
6. $\Gamma$ parameter in the Dynes function
7. Zero-bias conductance and resistivity
8. Calculation and summary of numerical values
9. Theoretical considerations of two-dimensional band structure and effective mass

Figs. S1 to S8

Table S1

References (*36-51*)

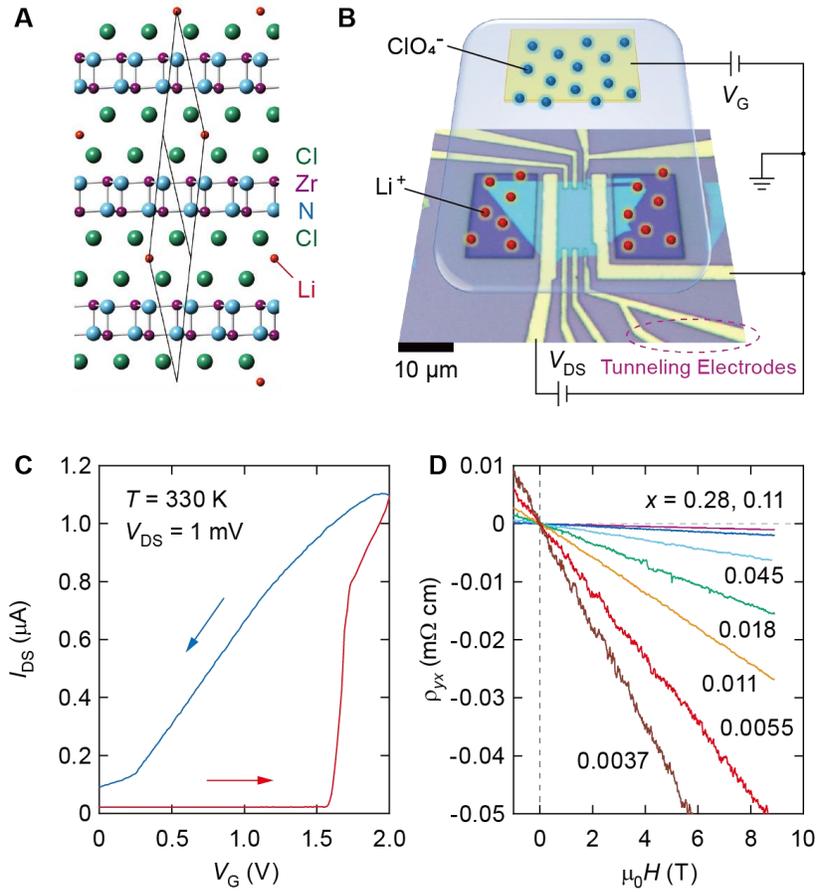

**Fig. 1. Gate controlled intercalation in ZrNCl device.** (**A**) Side view of Li$_x$ZrNCl crystal structure. Solid lines represent the rhombohedral unit cell. (**B**) Schematic illustration of the ionic-gating device based on a real optical micrograph picture of a ZrNCl single crystal flake and patterned electrodes. Gate voltage $V_G$ is applied to the electrolyte. Lithium ions intercalate from the sides of the covered flake. Narrow contacts are prepared for the tunneling spectroscopy measurements. (**C**) Source-drain current $I_{DS}$ of the device in intercalation operation. During the forward sweep of $V_G$ (red), $I_{DS}$ increases steeply, whereas the change of $I_{DS}$ is gradual in the backward scan (blue). $V_G$ is swept at a speed of 10 mV/sec. (**D**) Anti-symmetrized transverse resistivity at 150 K. The linear slope is used to determine the Li content $x$.

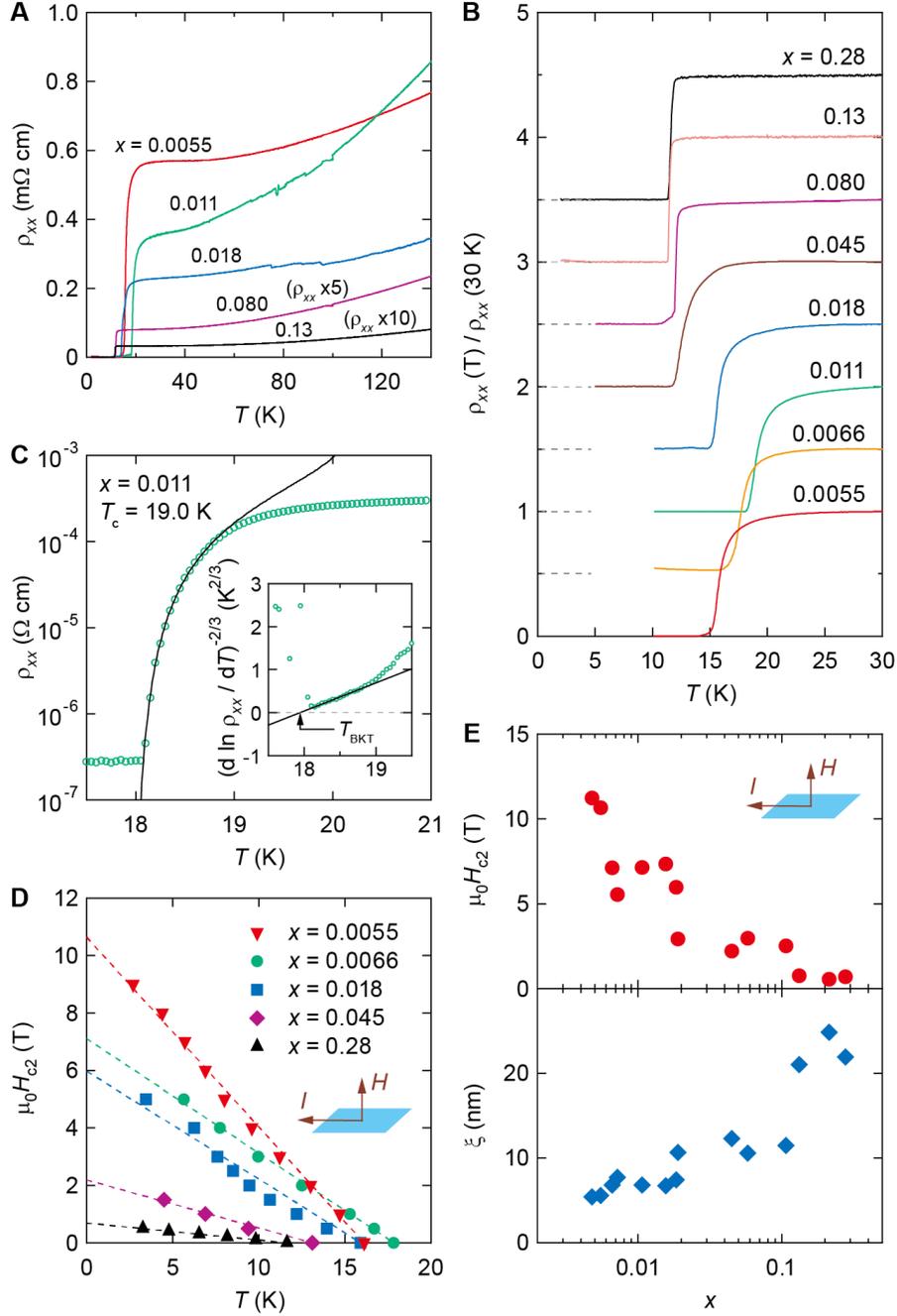

**Fig. 2. Transport properties of Li$_x$ZrNCl.** (**A**) Temperature dependence of resistivity at different doping levels. The resistivities at $x = 0.080$ and $0.13$ are multiplied by 5 and 10, respectively. (**B**) Resistivity normalized at 30 K. Each curve is shifted by 0.5, and gray dashed lines indicate zero lines. (**C**) Resistivity at $x = 0.011$ showing the BKT transition. The black line

is the fit to the Halperin-Nelson formula. Inset, resistivity plotted on a $[\mathrm{dln}(\rho)/\mathrm{d}T]^{-2/3}$ scale. (**D**) Out-of-plane upper critical field $H_{c2}$ as a function of temperature. Dashed lines are linear extrapolations to 0 K for each doping levels. (**E**) Doping dependence of $H_{c2}$ at 0 K in **D** (top) and in-plane coherence length $\xi$ (bottom).

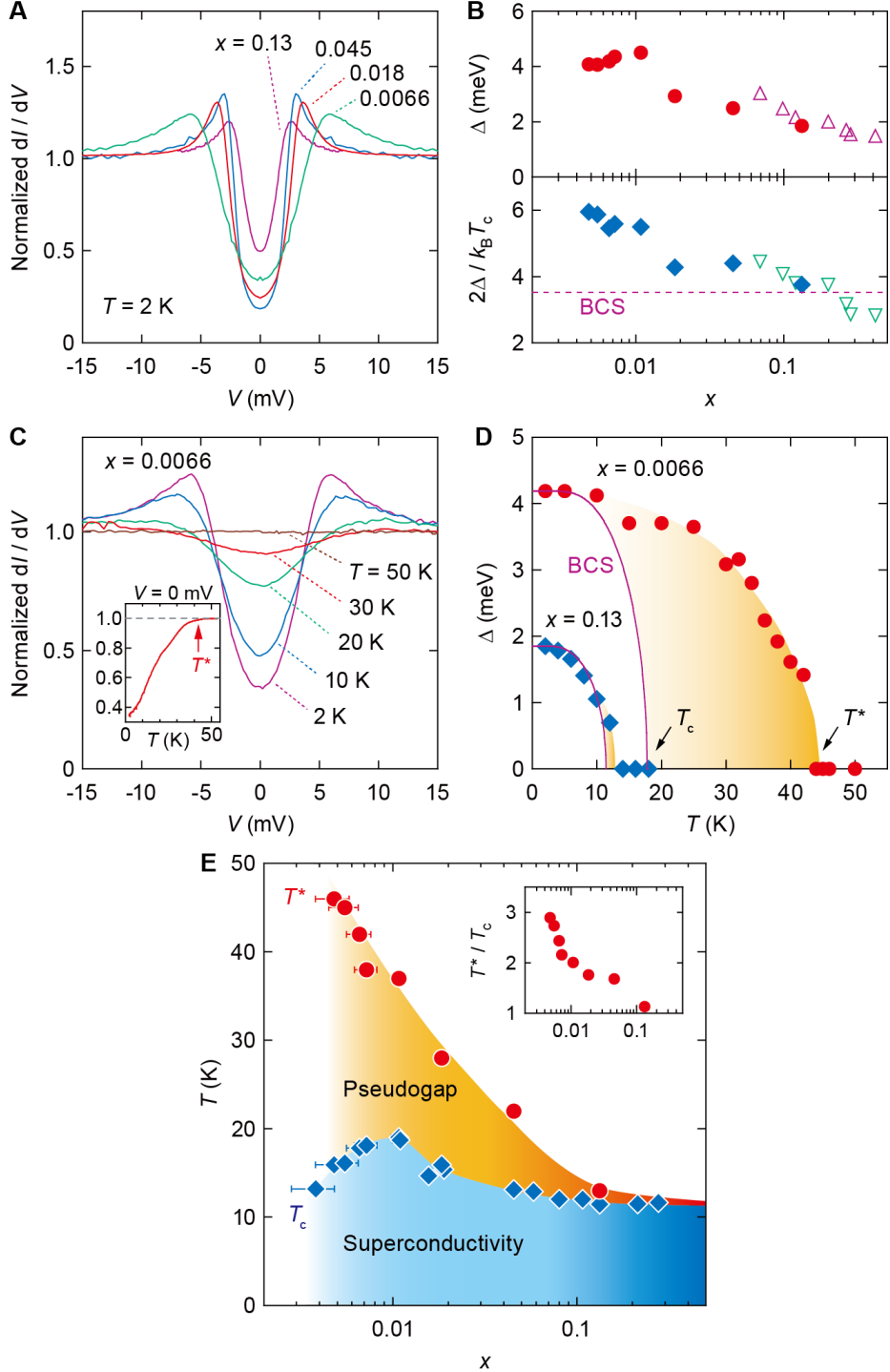

**Fig. 3. Tunneling spectroscopy of Li$_x$ZrNCl.** (**A**) Symmetrized tunneling spectra at 2 K. Each curve is normalized at 55 K after the subtraction of channel resistivity (*13*). (**B**) Doping dependence of superconducting gap Δ (top) and its ratio to the critical temperature $T_c$ (bottom).

The BCS theory predicts $2\Delta/k_\mathrm{B}T_\mathrm{c} = 3.52$ (dashed line). Open symbols are measured values in polycrystalline samples (*27*). (**C**) Tunneling spectra at $x = 0.0066$ without symmetrization for different temperatures. Inset, temperature scan of zero-bias-conductance (ZBC), $dI/dV$ at $V = 0$. Gap-opening temperature $T^*$ is determined by a 1% drop of ZBC. (**D**) $\Delta$ at $x = 0.0066$ (circles) and 0.13 (diamonds) as a function of temperature. Solid lines indicate the BCS-type gap function with $T_\mathrm{c}$ determined by the resistive transition. (**E**) Phase diagram of Li$_x$ZrNCl. The temperature regime between $T_\mathrm{c}$ and $T^*$ represents the pseudogap state. The error of carrier density is estimated by measurements in multiple Hall probes. Inset, the ratio between $T^*$ and $T_\mathrm{c}$.

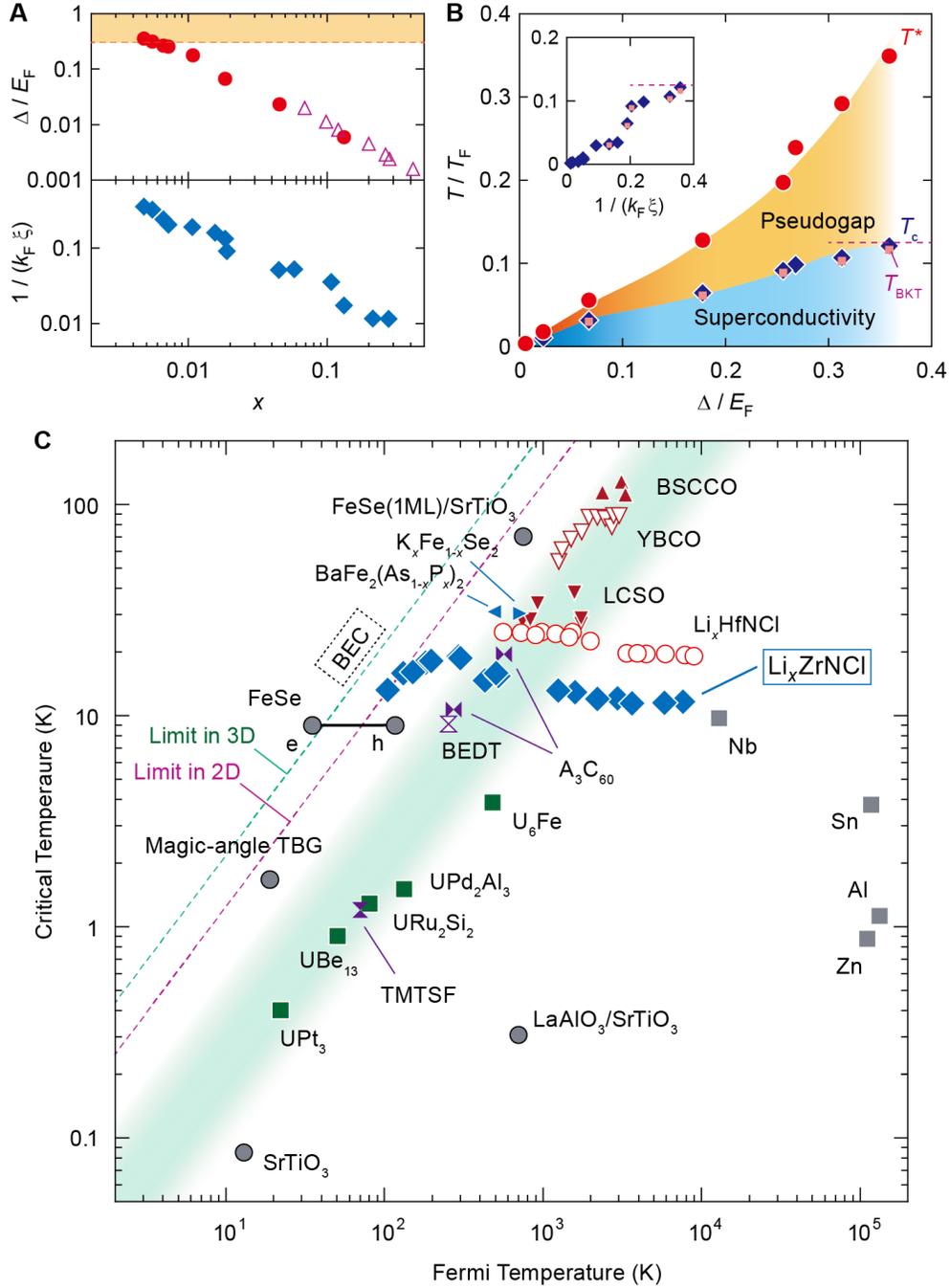

**Fig. 4. The BCS-BEC crossover in superconducting Li$_x$ZrNCl.** (**A**) Doping dependence of the ratio between superconducting gap and Fermi energy ($\Delta/E_F$) (top) and the ratio between interparticle distance and coherence length ($1/k_F\xi$) (bottom). The orange area represents the BCS-BEC crossover regime (*20*). Open triangles are measured values by the specific heat

measurement (*27*). (**B**) The phase diagram of the BCS-BEC crossover. Gap opening temperature $T^*$, critical temperature $T_c$ and critical temperature of BKT transition $T_{BKT}$ are normalized by Fermi temperature $T_F$ and plotted as functions of $\Delta/E_F$ with red spheres, dark blue diamonds, and pink squares, respectively. The dashed line represents the theoretically predicted upper bound, $T_{BKT}/T_F = 0.125$ (*22*). Inset, $T_c/T_F$ and $T_{BKT}/T_F$ as functions of $1/k_F\xi$. (**C**) Uemura plot: Critical temperature *vs.* Fermi temperature is plotted for various superconductors (*11-13, 28-30*). Li$_x$ZrNCl traverses from BCS to the crossover region across the shaded area, where most of the unconventional superconductors are located (*8*). The dashed line denoted as "Limit in 3D" represents the critical temperature in the BEC limit in 3D Fermi gas systems, $T_c = 0.218\ T_F$ (*2*). The other dashed line, denoted as "Limit in 2D", corresponds to the general upper limit of $T_{BKT} = 0.125\ T_F$ in all 2D fermionic systems (*22*).

# Supplementary Materials for

## Gate-controlled BCS-BEC crossover in a two-dimensional superconductor

Yuji Nakagawa, Yuichi Kasahara, Takuya Nomoto, Ryotaro Arita,
Tsutomu Nojima, Yoshihiro Iwasa*

Correspondence to: iwasa@ap.t.u-tokyo.ac.jp

**This PDF file includes:**

Materials and Methods
    Device fabrication
    Measurement details

Supplementary Text
1. Low carrier density superconductors
2. Temperature dependence of Hall coefficient
3. Uniformity of intercalation
4. Measurement of the upper critical field
5. Raw data and symmetrization of tunneling spectra
6. $\Gamma$ parameter in the Dynes function
7. Zero-bias conductance and resistivity
8. Calculation and summary of numerical values
9. Theoretical considerations of two-dimensional band structure and effective mass

Figs. S1 to S8
Table S1
References (*36-51*)

**Materials and Methods**

Device fabrication

Pristine ZrNCl powders were prepared by a chemical vapor transport method (*36*). We isolated single-crystalline flakes onto a SiO$_2$/Si substrate with the Scotch-tape exfoliation and transfer. The thicknesses of the selected flakes were in the range of tens of nanometers as determined by atomic force microscopy. By the standard electron-beam lithography process with PMMA as a resist, electrodes including a side gate were patterned, and Ti (7 nm) and Au (70 nm) were deposited. We covered Hall-bar electrodes and the channel with PMMA, whereas the side gate and the flake outside the channel region were exposed. We dissolved LiClO$_4$ (Sigma Aldrich) in polyethylene glycol (PEG; $M_\text{w}$ = 600, Wako) at 80°C under vacuum. The ratio between Li and O in PEG was set to be 1:20. We put a droplet of the electrolyte solution to cover both the sample and the gate electrode and subsequently transferred the device into a Quantum Design Physical Property Measurement System. Before starting measurements, the device was kept in vacuum using the high-vacuum option (at a pressure below $10^{-4}$ Torr) at 330 K for more than 1 hour.

Measurement details

We measured the resistance of the device with a standard four-probe geometry. The function generator (NF Corporation WF1974) input ac voltage, and lock-in amplifiers (Stanford Research Systems Model SR830 DSP and Signal Recovery Model 5210) measured current and voltage. The gate voltage was applied at 330 K under vacuum using a Keithley 2400 source meter. After cooling down to 150 K, which is lower than the melting temperature of PEG (288 K), the chamber was purged with He, and we measured the Hall coefficient. Tunneling spectroscopy was performed with ac and dc excitations. The ac and dc voltages were probed with the lock-in amplifier and the dc voltmeter (Keithley Model 2182A), respectively.

**Supplementary Text**

1. <u>Low carrier density superconductors</u>

As well as the exotic superconductors, the low-carrier density superconductors are also candidates to realize the crossover because of their small $E_F$. However, their $T_c$ and $\Delta$ tend to be small (*13, 37, 38*). Moreover, it usually decreases when we reduced the carrier density. These features are represented by SrTiO$_3$. The substantial values of $T_c/T_F$ and $\Delta/E_F$ are still challenging to obtain.

Li$_x$ZrNCl, however, sustains high $T_c$ in the low carrier density regime, as shown in Fig. S1. At $n = 10^{20}$ cm$^{-3}$, SrTiO$_3$ shows the maximum $T_c = 480$ mK, whereas $T_c$ of Li$_x$ZrNCl is 16 K. The high-$T_c$ low carrier density superconductivity is crucial to realize the BCS-BEC crossover.

2. <u>Temperature dependence of Hall coefficient</u>

Figure S2 shows a typical result of Hall measurements at different temperatures. From 150 K to 50 K, which is just above the gap opening temperature $T^*$, the Hall coefficient is almost identical. This is consistent with the Fermi liquid behavior and the previous report for polycrystalline samples (*39*). In this study, we adopted the value at 150 K to calculate the carrier density and the doping level.

3. <u>Uniformity of intercalation</u>

We confirmed the uniformity of a device after intercalation by using different sets of electrodes. Figures S3A and S3B show longitudinal and transverse resistivity, respectively, of a typical device. The value of $\rho_{xx}$, the superconducting transition, and Li concentration determined with the Hall coefficient are almost the same, regardless of probes, representing the uniform intercalation in our micro-scale single crystal device. In Fig. 3E in the main text, the small positional dependence of Li concentration is represented by error bars. We also note that the superconducting transitions are smooth without multi-step transitions (see Fig. 2B), indicating that there is no phase separation into different doping levels with different $T_c$s.

4. Measurement of the upper critical field

We determined the upper critical field $H_{c2}$ by measuring the resistivity under magnetic fields (Fig. S4). The transition point is defined as follows. The resistivity at high temperature (>30 K) under out-of-plane magnetic field of 9 T was extrapolated using a polynomial function of $T$, $\rho_{xx}(T) = a + bT + cT^2$. We used this curve as the normal state resistance. The crossing point between the half value of the extrapolated curve and the measured curve at each magnetic field is plotted in Fig. 2D.

5. Raw data and symmetrization of tunneling spectra

Fig. S5A shows the raw data of the tunneling spectra at $x = 0.045$. The data obtained above $T_c$ include the contribution of the channel resistance, which should be subtracted, as explained in our previous work (*13*). The channel resistance was estimated from the value of $dI/dV$ far outside the superconducting gap because they are expected to be temperature independent. Usually, this estimation was conducted without symmetrization (for example, Fig. 3C). However, when the estimated channel resistance is different between the positive and negative biases, we symmetrized the spectra (Fig. S5B) as $dI/dV$ (symmetrized)$(V) = (dI/dV (+V) + dI/dV (-V))/2$, to obtain bias-independent channel resistance.

6. Γ parameter in the Dynes function

After the normalization process of the tunneling spectra (*13*), we obtained the superconducting gap Δ by the fitting using the Dynes function (*26*) as the energy-dependent density of states ($N(E)$).

$$N(E) = \left| \mathrm{Re} \left( \frac{E - i\Gamma}{\sqrt{(E - i\Gamma)^2 - \Delta^2}} \right) \right|, \quad (S1)$$

where Γ is the broadening factor. The temperature and doping dependences of Δ and Γ are plotted in Fig. S6. The larger value of Γ at high temperatures and in the low carrier density regime is consistent with other tunneling studies (*13, 34*). It is noted that Γ exhibits sudden increase above 20 K for $x = 0.0066$, which possibly signifies the decoherence above $T_c$.

7. Zero-bias conductance and resistivity

The gap opening temperature $T^*$ was determined by $T$ dependence of zero-bias-conductance (ZBC), which is $dI/dV$ at $V = 0$ at the tunneling junction. ZBC was normalized in the same process as the tunneling spectra. We defined $T^*$ as the temperature where ZBC decreases by 1% (Fig. S7). It is noted that $T^*$ was not detected in the longitudinal resistivity. In the low-doping levels, $T^*$ is apart from the superconducting transition temperature $T_c$ defined from the resistive transition.

8. Calculation and summary of numerical values

In Table S1, we summarized obtained values of carrier density $n$ at 150 K, Fermi wave vector $k_F$, Fermi energy $E_F$, superconducting gap Δ at 2 K, critical temperature $T_c$, gap-opening temperature $T^*$, the out-of-plane upper critical field $\mu_0 H_{c2}$ linearly extrapolated to 0 K, in-plane coherence length $\xi$ at 0 K, Pippard coherence length $\xi_{\text{Pippard}}$, and the mean free path at 50 K for various $x$ in Li$_x$ZrNCl.

$k_F$ and $E_F$ were calculated from $n$ by considering an ideal parabolic band dispersion in two-dimension. $k_F = (4\pi n_{\text{layer}}/ss')^{1/2}$ and $E_F = \hbar^2 k_F^2/2m^*$, where $n_{\text{layer}}$ is two-dimensional carrier density per layer, $s$ is the spin degree of freedom, $s'$ is the valley degree of freedom, $\hbar$ is the Dirac's constant, and $m^*$ is the effective mass of electrons. For Li$_x$ZrNCl, $s = s' = 2$. We used $m^* = 0.9 m_0$, as reported in a previous experiment (27). $\xi_{\text{Pippard}}$ was calculated as $\hbar v_F/\pi\Delta$, where $v_F$ is the Fermi wave velocity $\hbar k_F/m^*$.

9. <u>Theoretical considerations of two-dimensional band structure and effective mass</u>

We have redone the density functional theory (DFT) calculation of Li$_x$ZrNCl in the low carrier density regime and with higher energy resolutions. Figures S8A and S8B show the band structures of ZrNCl. Note that the energy scales of in-plane (Fig. S8A) and out-of-plane (Fig. S8B) are more than two orders of magnitude different. Even so, the dispersion of the conduction band bottom is indiscernible (<0.1 meV) along the *K-H* line. The absence of *K-H* dispersion even at the high resolution level is fairly consistent with the symmetry consideration of the band structure of the rhombohedral lattice, where the interlayer hopping vanishes up to the second nearest layers (*25, 40*), as mentioned in the main text. This indicates that bulk Li$_x$ZrNCl is an ideal 2D system in the low doping level. Figure S8C shows the density of states against chemical potential $\mu$ measured from the conduction band edge. The density of states exhibits an extremely sharp jump at around $\mu = 0$ followed by the flat spectrum, also supporting the high two-dimensionality.

The effective mass of the conduction band bottom is obtained as $m_{DFT}^* \sim 0.58 m_0$ from the least square fit in the energy region of 0 eV $< \varepsilon_k <$ 0.2 eV. This is consistent with the previous calculations (*17, 41, 42*). This effective mass is slightly smaller than the experimental value of $m^* = 0.9\ m_0$ (*27*), but is known to be enhanced by the electron-phonon coupling.

About the electron-phonon interactions, we calculate the renormalization factor $Z(\varepsilon_k)$ based on density functional theory for superconductors (SCDFT) (*43, 44*) and obtain the dimensionless electron-phonon coupling strength λ by taking an average of $Z(\varepsilon_k)$ on the Fermi surface. We overlay the μ dependence of λ in Fig. S8C. We can see that λ~0.6 for μ~10-20 meV, corresponding to the low-doped samples ($x$~0.005-0.01). These values are consistent with the previous calculations made in the higher doping regimes $x > 0.05$ (*17, 45*). Since $m^*$ is enhanced by a factor of $1 + \lambda$, we obtain $m^* \sim 0.93 m_0$, which shows fair agreement with the experiment.

The actual calculations are carried out as follows: First, we perform an electronic structure calculation based on DFT with Quantum Espresso code (*46, 47*). The lattice constants and the internal coordinates of atoms are taken from the experiment. We employ the non-relativistic version of Optimized Norm-Conserving Vanderbilt pseudopotentials (*48, 49*), the exchange-correlation functional proposed by Perdew *et al.* (*50*), and $10^3$ ***k***-mesh for the self-consistent calculation. The cutoff energy of the plane waves is set to 90 Ry. After the electronic structure calculation, we perform a phonon frequency and electron-phonon vertex calculations based on density functional perturbation theory (DFPT) (*51*). Here, $5^3$ ***q***-mesh for the phonon momentum is employed. Finally, we calculate the renormalization factor $Z(\varepsilon_k)$ based on density functional theory for superconductors and obtain λ as an average of $Z(\varepsilon_k)$ on the Fermi surface. About the chemical potential μ dependence of λ, we simply shift μ in the calculation of $Z(\varepsilon_k)$ keeping the electron-phonon vertex in the non-doped system. The calculations of λ as well as the density of states are carried out with dense $\mathbf{k}$-meshes $N_k = 80^3$.

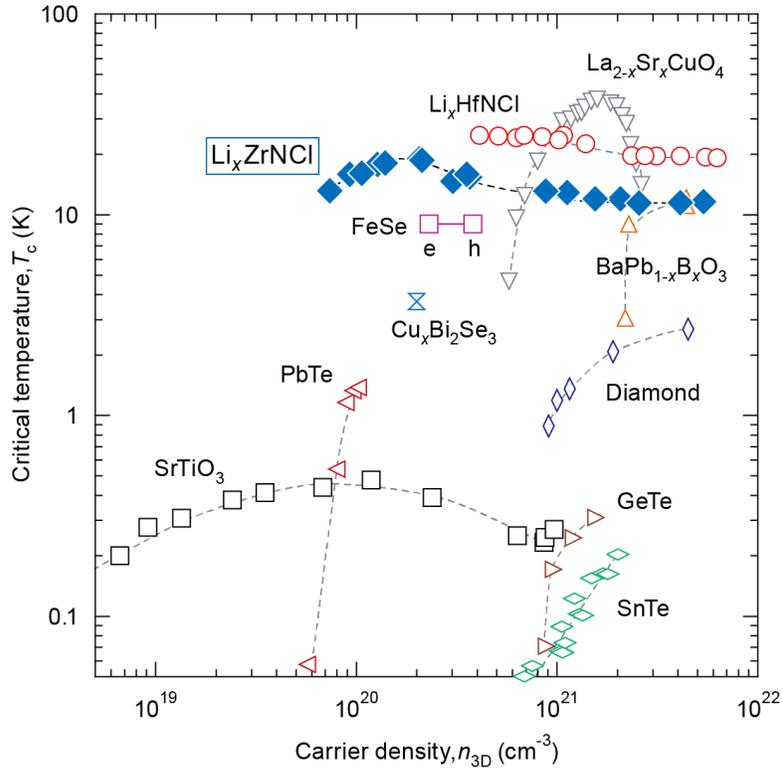

**Fig. S1. Low carrier density superconductors.**

Logarithmic plot of critical temperature $T_c$ vs. 3D carrier density $n_{3D}$. $T_c$ of Li$_x$ZrNCl remains high at $n_{3D} = 10^{20}$ cm$^{-3}$, where superconductivity in other systems disappears or has a small $T_c$ (*13, 37, 38*).

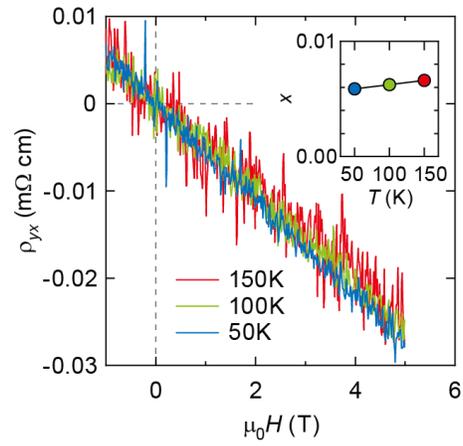

**Fig. S2. Temperature dependence of Hall measurements.**

Anti-symmetrized transverse resistivity as a function of magnetic field at 50, 100, and 150 K. The calculated doping levels are plotted in the inset, which shows small temperature dependence. We used the value at 150 K in the main text.

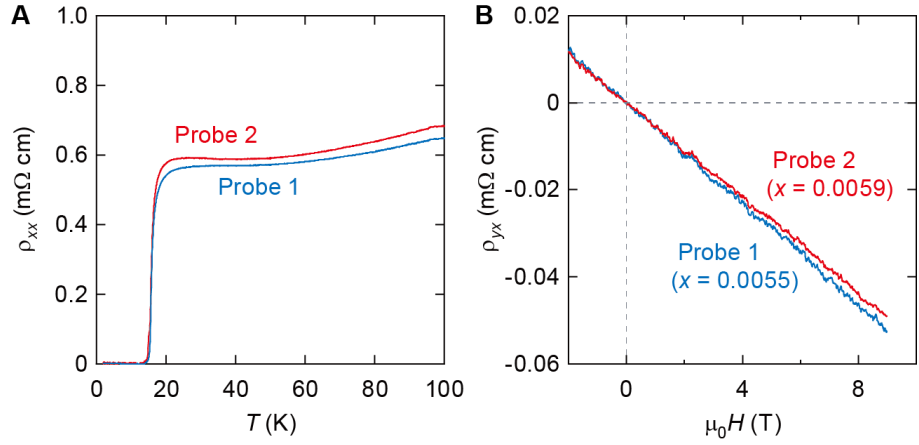

**Fig. S3. Positional dependence of resistivity.**

(**A**) Temperature dependence of resistivity measured with different sets of four-terminal voltage probes in the same device. The resistivity values and superconducting transitions are almost identical. (**B**) Anti-symmetrized transverse resistivity as a function of magnetic field measured with different sets of Hall bar probes at 150 K. The little difference of calculated doping levels is shown in the error bars in Fig. 3E.

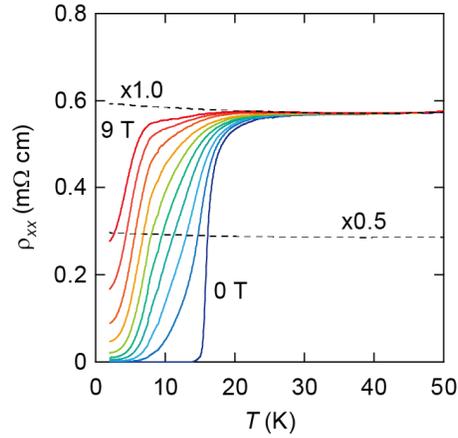

**Fig. S4. Resistivity under out-of-plane magnetic fields.**

Solid lines are temperature-dependent resistivity under out-of-plane magnetic field of 0, 1, …, 9 T. The upper dashed line is the extrapolation of the resistivity under 9 T at $T > 30$ K and used as normal state resistance. The $H_{c2}$-$T$ plot (Fig. 2D) was made from the crossing points between solid lines and the lower dashed line, which is the half of the normal state resistance.

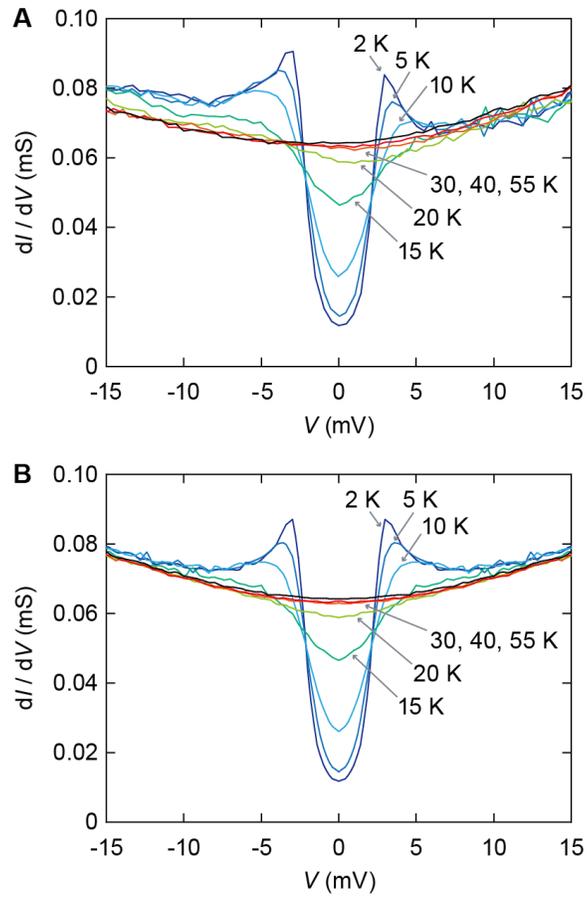

**Fig. S5. Raw and symmetrized tunneling spectra.**

(**A**) Raw tunneling spectra measured at different temperatures at $x = 0.045$. (**B**) The spectra after symmetrization. The difference between low and high temperatures at high-voltage regime is used for estimating channel resistance's contribution (*13*).

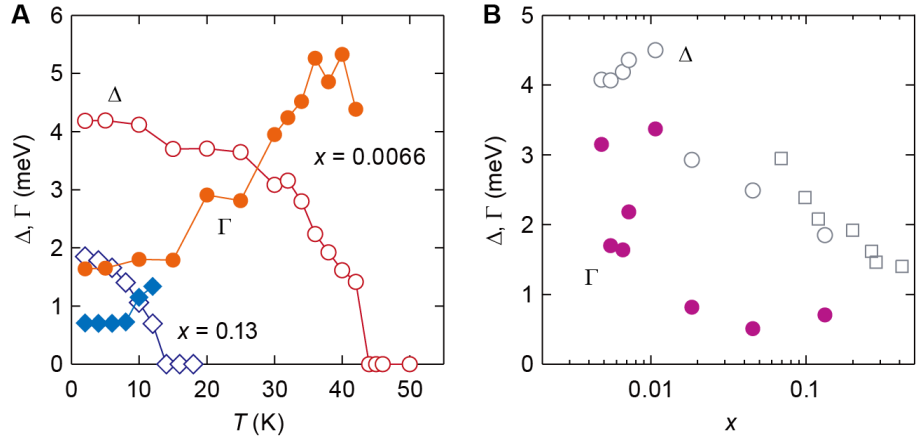

**Fig. S6. Fitting results of the tunneling spectra.**

(**A**) Temperature dependence of the superconducting gap Δ (open symbols) and the broadening parameter Γ (filled symbols) at $x = 0.13$ (blue) and 0.0066 (red). (**B**) Doping dependence of Δ (open symbols) and Γ (filled symbols) at 2 K. The square symbols of Δ represents the results of the specific heat measurements (*27*). The points of Δ are identical to Fig. 3D in the main text.

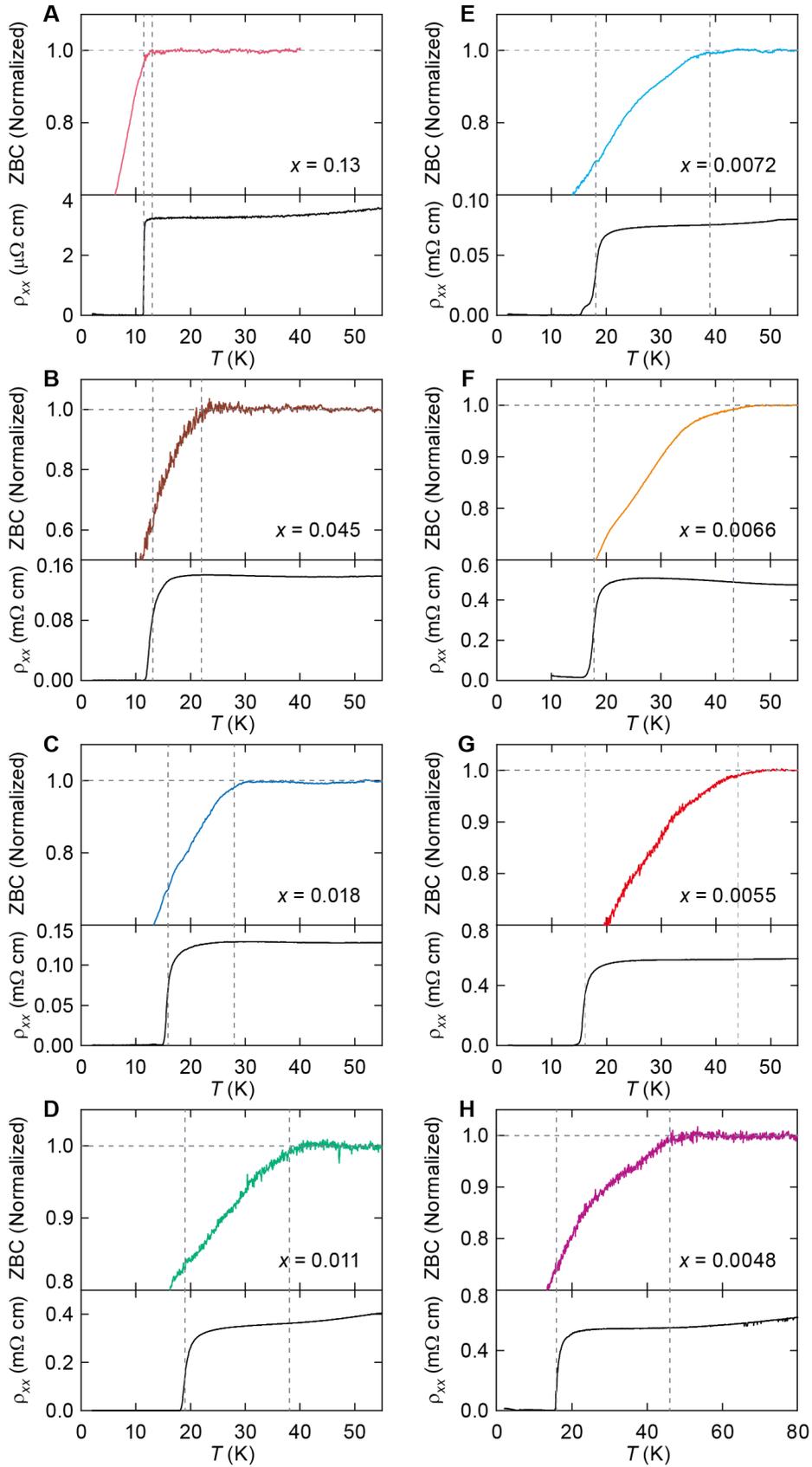

**Fig. S7. Zero-bias-conductance and resistivity**

Temperature dependence of normalized zero-bias-conductance (ZBC, top) and resistivity ($\rho_{xx}$, bottom) are plotted for various doping levels. The horizontal dashed lines correspond to ZBC = 1. The vertical dashed lines indicate $T = T_c$ (the midpoint of the transition in $\rho_{xx}$) and $T^*$ (the point where ZBC decreases by 1%).

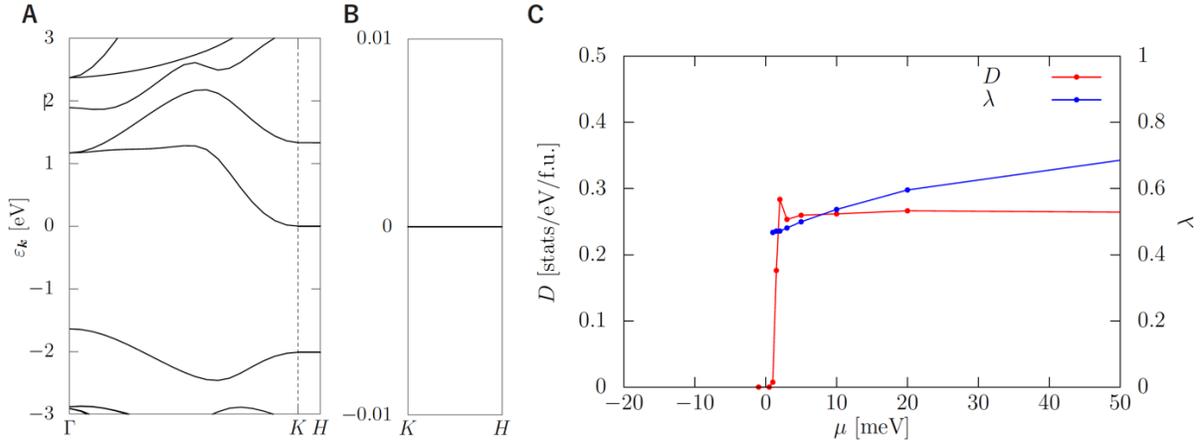

**Fig. S8. Band structure, density of states, and electron-phonon coupling constant**

(A) The band structure of ZrNCl, where $\varepsilon_k = 0$ corresponds to the energy of the conduction band bottom. The symbols for high symmetry points follow the notation in the hexagonal Brillouin zone although the primitive cell of ZrNCl is rhombohedral. (B) The enlarged band structure along the *K-H* line. (C) The density of states $D$ (red) and the dimensionless electron-phonon coupling constant $\lambda$ (blue) as functions of the chemical potential $\mu$. The energy of the conduction band bottom is set to $\mu=0$. These calculations were made with the dense ***k***-mesh $N_k = 80^3$.

**Table S1. Summary of doping dependence in Li$_x$ZrNCl.**

| $x$ | $n$ ($\times 10^{20}$ cm$^{-3}$) | $E_F$ (meV) | $k_F$ (nm$^{-1}$) | $T_c$ (K) | $T^*$ (K) | $\Delta$ (meV) |
|---|---|---|---|---|---|---|
| 0.00480 | 0.927 | 11.3 | 0.518 | 15.9 | 46 | 4.08 |
| 0.00550 | 1.06 | 13.0 | 0.554 | 16.1 | 44 | 4.07 |
| 0.00659 | 1.27 | 15.6 | 0.607 | 17.8 | 43 | 4.19 |
| 0.00720 | 1.39 | 17.0 | 0.634 | 18.1 | 39 | 4.36 |
| 0.0107 | 2.07 | 25.3 | 0.773 | 19.0 | 38 | 4.50 |
| 0.0184 | 3.55 | 43.5 | 1.01 | 15.9 | 28 | 2.93 |
| 0.0454 | 8.77 | 107 | 1.59 | 13.1 | 22 | 2.49 |
| 0.133 | 25.6 | 313 | 2.72 | 11.5 | 13 | 1.85 |

| $x$ | $\mu_0 H_{c2}$ (T) | $\xi$ (nm) | $\xi_{Pippard}$ (nm) | mean free path (nm) |
|---|---|---|---|---|
| 0.00480 | 11.2 | 5.41 | 3.42 | 12.4 |
| 0.00550 | 10.6 | 5.56 | 3.67 | 3.55 |
| 0.00659 | 7.11 | 6.80 | 3.90 | 3.97 |
| 0.00720 | 5.54 | 7.71 | 3.92 | 27.9 |
| 0.0107 | 7.13 | 6.80 | 4.63 | 3.92 |
| 0.0184 | 5.97 | 7.42 | 9.31 | 4.63 |
| 0.0454 | 2.19 | 12.3 | 17.3 | 17.8 |
| 0.133 | 0.746 | 21.0 | 39.6 | 127 |

**References**

52. S. Yamanaka, H. Kawaji, K. -i. Hotehama, M. Ohashi, A new layer-structured nitride superconductor. Lithium-intercalated β-zirconium nitride chloride, Li$_x$ZrNCl. *Adv. Mater.* **8**, 771-774 (1996).

53. J. K. Hulm, C. K. Jones, D. W. Deis, H. A. Fairbank, P. A. Lawless, Superconducting Interactions in Tin Telluride. *Phys. Rev.* **169**, 388-394 (1968).

54. T. D. Thanh, A. Koma, S. Tanaka, Superconductivity in the BaPb$_{1-x}$Bi$_x$O$_3$ system. *Appl. Phys.* **22**, 205-212 (1980).

55. T. Takano, A. Kitora, Y. Taguchi, Y. Iwasa, Modulation-doped-semiconductorlike behavior manifested in magnetotransport measurements of Li$_x$ZrNCl layered superconductors. *Phys. Rev. B* **77**, 104518 (2008).

56. R. Akashi, M. Ochi, R. Suzuki, S. Bordács, Y. Tokura, Y. Iwasa, R. Arita, Two-Dimensional Valley Electrons and Excitons in Noncentrosymmetric 3*R*-MoS$_2$, *Phys. Rev. Appl.* **4**, 014002 (2015).

57. R. Weht, A. Filippetti, W. E. Pickett, Electron doping in the honeycomb bilayer superconductors (Zr, Hf) NCl. *Europhys. Lett.* **48**, 320-325 (1999).

58. R. Heid, K. P. Bohnen, *Ab Initio* lattice dynamics and electron-phonon coupling in Li$_x$ZrNCl. *Phys. Rev. B* **72**, 134527 (2005).

59. M. Lüders, M. A. L. Marques, N. N. Lathiotakis, A. Floris, G. Profeta, L. Fast, A. Continenza, S. Massidda, E. K. U. Gross, *Phys. Rev. B* **72**, 024545 (2005).

60. M. Kawamura, R. Akashi, S. Tsuneyuki, Anisotropic superconducting gaps in YNi$_2$B$_2$C: A first-principles investigation. *Phys. Rev. B* **95**, 054506 (2017).

61. R. Akashi, K. Nakamura, R. Arita, M. Imada, High-temperature superconductivity in layered nitrides β-Li$_x$$M$NCl ($M$ = Ti, Zr, Hf): Insights from density functional theory for superconductors. *Phys. Rev. B* **86**, 054513 (2012).

62. P. Giannozzi, S. Baroni, N. Bonini, M. Calandra, R. Car, C. Cavazzoni, D. Ceresoli, G. L. Chiarotti, M. Cococcioni, I. Dabo, A. Dal Corso, S. de Gironcoli, S. Fabris, G. Fratesi, R. Gebauer, U. Gerstmann, C. Gougoussis, A. Kokalj, M. Lazzeri, L. Martin-Samos, N. Marzari, F. Mauri, R. Mazzarello, S. Paolini, A. Pasquarello, L. Paulatto, C. Sbraccia, S. Scandolo, G. Sclauzero, A. P. Seitsonen, A. Smogunov, P. Umari, R. M. Wentzcovitch, QUANTUM ESPRESSO: a modular and open-source software project for quantum simulations of materials. *J. Phys.: Condens. Matter* **21**, 395502 (2009).

63. P. Giannozzi, O. Andreussi, T. Brumme, O. Bunau, M. B. Nardelli, M. Calandra, R. Car, C. Cavazzoni, D. Ceresoli, M. Cococcioni, N. Colonna, I. Carnimeo, A. Dal Corso, S. de Gironcoli, P. Delugas, R. A. DiStasio Jr., A. Ferretti, A. Floris, G. Fratesi, G. Fugallo, R. Gebauer, U. Gerstmann, F. Giustino, T. Gorni, J. Jia, M. Kawamura, H. -Y. Ko, A. Kokalj, E. Küçükbenli, M. Lazzeri, M. Marsili, N. Marzari, F. Mauri, N. L. Nguyen, H. -V. Nguyen, A. Otero-de-la-Roza, L. Paulatto, S. Poncé, D. Rocca, R. Sabatini, B. Santra, M. Schlipf, A. P. Seitsonen, A. Smogunov, I. Timrov, T. Thonhauser, P. Umari, N. Vast, X. Wu, S. Baroni Advanced capabilities for materials modelling with Quantum ESPRESSO. *J. Phys.: Condens. Matter* **29**, 465901 (2017).

64. D. R. Hamann, Optimized norm-conserving Vanderbilt pseudopotentials. *Phys. Rev. B* **88**, 085117 (2013).

65. P. Scherpelz, M. Govoni, I. Hamada, G. Galli, Implementation and validation of fully relativistic *GW* calculations: Spin–orbit coupling in molecules, nanocrystals, and solids. *J. Chem. Theory Comput.* **12**, 3523-3544 (2016).

66. J. P. Perdew, K. Burke, M. Ernzerhof, Generalized gradient approximation made simple. *Phys. Rev. Lett.* **77**, 3865-3868 (1996).

67. S. Baroni, S. de Gironcoli, A. Dal Corso, P. Giannozzi, Phonons and related crystal properties from density-functional perturbation theory. *Rev. Mod. Phys.* **73**, 515-562 (2001).